\documentclass[pra,twocolumn,nofootnoteinbib]{revtex4}

\usepackage[verbose=true]{microtype}
\usepackage{amsmath}
\usepackage{amssymb}
\usepackage{txfonts}
\usepackage{mathtools}
\usepackage{color}
\usepackage[colorlinks,linkcolor=black,citecolor=black,linktocpage,hypertexnames=true,pdfpagelabels]{hyperref}
\usepackage[amsmath,thmmarks,hyperref]{ntheorem}
\usepackage{amsmath}
\usepackage[latin1]{inputenc}
\usepackage[capitalise]{cleveref}
\usepackage[all]{xy}
\usepackage{array}
\usepackage{enumerate}
\usepackage{enumitem}
\usepackage{tikz}
\usepackage{relsize}
\usepackage[capitalise]{cleveref}
\usepackage[textwidth=.35in,textsize=scriptsize,colorinlistoftodos]{todonotes}
\usepackage{braket}
\usepackage{float}
\usepackage{graphicx}

\newtheorem{theorem}{Theorem}

\newtheorem{definition}{Definition}
\newtheorem{proposition}{Proposition}

\newtheorem*{proposition-nn}{Proposition}

\theorembodyfont{\normalfont}
\theoremsymbol{$\Box$}

\theorembodyfont{\normalfont}
\theoremsymbol{$\Box$}
\newtheorem{example}{Example}

\theorembodyfont{\normalfont}
\theoremsymbol{$\Box$}

\theoremstyle{nonumberplain}
\theorembodyfont{\normalfont}
\theoremsymbol{$\Box$}
\newtheorem{proof}{Proof}

\renewcommand{\epsilon}{\varepsilon}
\newcommand{\E}{\mathcal{E}}
\newcommand{\M}{\mathcal{M}}

\newcommand{\N}{\mathbb{N}}

\renewcommand{\L}{\mathcal{L}}

\newcommand{\be}{\begin{eqnarray}}
\newcommand{\ee}{\end{eqnarray}}
\newcommand{\ben}{\begin{enumerate}}
\newcommand{\een}{\end{enumerate}}
\newcommand{\im}{\item}
\newcommand{\ba}{\begin{array}}
\newcommand{\ea}{\end{array}}

\newcommand{\ra}{\rangle}
\newcommand{\la}{\langle}
\newcommand{\mc}{\mathcal}

\definecolor{Gray}{gray}{0.3}
\definecolor{LightCyan}{rgb}{0.88,1,1}
\definecolor{orange}{rgb}{1,0.5,0}
\definecolor{amber}{rgb}{1.0, 0.75, 0.0}
\definecolor{amaranth}{rgb}{0.9, 0.17, 0.31}
\definecolor{antiquefuchsia}{rgb}{0.57, 0.36, 0.51}
\definecolor{antiquewhite}{rgb}{0.98, 0.92, 0.84}
\definecolor{arylideyellow}{rgb}{0.91, 0.84, 0.42}
\definecolor{babyblue}{rgb}{0.54, 0.81, 0.94}
\definecolor{babypink}{rgb}{0.96, 0.76, 0.76}
\definecolor{bittersweet}{rgb}{1.0, 0.44, 0.37}
\definecolor{bleudefrance}{rgb}{0.19, 0.55, 0.91}
\definecolor{blue(munsell)}{rgb}{0.0, 0.5, 0.69}
\definecolor{blue(pigment)}{rgb}{0.2, 0.2, 0.6}
\definecolor{blue-violet}{rgb}{0.54, 0.17, 0.89}
\definecolor{bananamania}{rgb}{0.98, 0.91, 0.71}
\definecolor{celadon}{rgb}{0.67, 0.88, 0.69}
\definecolor{cherryblossompink}{rgb}{1.0, 0.72, 0.77}
\definecolor{azure(colorwheel)}{rgb}{0.0, 0.5, 1.0}
\definecolor{darkpink}{rgb}{0.91, 0.33, 0.5}
\definecolor{darkraspberry}{rgb}{0.53, 0.15, 0.34}
\definecolor{flame}{rgb}{0.89, 0.35, 0.13}
\definecolor{lightcarminepink}{rgb}{0.9, 0.4, 0.38}
\definecolor{mediumred-violet}{rgb}{0.73, 0.2, 0.52}
\definecolor{mediumtealblue}{rgb}{0.0, 0.33, 0.71}
\definecolor{mediumturquoise}{rgb}{0.28, 0.82, 0.8}
\definecolor{darkmagenta}{rgb}{0.55, 0.0, 0.55}
\definecolor{chromeyellow}{rgb}{1.0, 0.65, 0.0}
\definecolor{flame}{rgb}{0.89, 0.35, 0.13}
\definecolor{darktangerine}{rgb}{1.0, 0.66, 0.07}
\definecolor{airforceblue}{rgb}{0.36, 0.54, 0.66}
\definecolor{amethyst}{rgb}{0.6, 0.4, 0.8}
\definecolor{ao}{rgb}{0.0, 0.5, 0.0}
\definecolor{arsenic}{rgb}{0.23, 0.27, 0.29}
\definecolor{Gray}{gray}{0.3}
\definecolor{LightCyan}{rgb}{0.88,1,1}
\definecolor{darkblue}{rgb}{0.0, 0.0, 0.55}

\newcommand{\nn}{\nonumber}

\makeatletter
\def\p@subsection{}
\def\p@subsubsection{}
\makeatother

\newcommand{\pp}{\mathcal{P}}

\begin{document}

\title{Generalised ansatz for continuous Matrix Product States}
\author{Maria Balanzó-Juandó and Gemma De las Cuevas}
\affiliation{Institut für Theoretische Physik, Universität Innsbruck, Technikerstr.\ 21a, 6020 Innsbruck, Austria}

\begin{abstract}
Recently it was shown that continuous Matrix Product States (cMPS) cannot express the continuum limit state of any Matrix Product State (MPS), according to a certain natural definition of the latter. The missing element is a projector in the transfer matrix of the MPS. Here we provide a generalised ansatz of cMPS that is capable of expressing the continuum limit of any MPS. It consists of a sum of cMPS with different boundary conditions, each attached to an ancilla state. This new ansatz can be interpreted as the concatenation of a state which is at the closure of the set of cMPS together with a standard cMPS. The former can be seen as a cMPS in the thermodynamic limit, or with matrices of unbounded norm. We provide several examples and discuss the result. 
\end{abstract}

\maketitle

\section{Introduction}

Tensor networks are a powerful ansatz to describe and simulate quantum many-body systems in an efficient way \cite{Or14b,Or18}, which originated in the context of condensed matter physics, but has nowadays percolated to other areas---for a recent review see \cite{Or18}. 
 The class of Matrix Product States (MPS) has been extremely successful in 
capturing ground states of gapped Hamiltonians in one spatial dimension \cite{Fa92,Vi03,Sc11d,Pe07}, and many generalisations to different settings have been considered: 
 to higher dimensional lattices \cite{Ve04a}, 
 to describe mixed states \cite{Ve04d,Zw04}, 
 to describe fermionic systems \cite{Kr10b}, 
in the context of conformal field theories \cite{Ci09}, 
or to describe ground states of critical systems \cite{Vi07}, 
to cite a few. 
The generalisation that is relevant for this work is the one to continuous systems, that is, where the lattice is replaced by a segment of the reals $\mc{R}=[x,y]$, but the bond dimension remains finite. 
Specifically, continuous MPS (cMPS) were proposed in 2010 as a tensor network ansatz to describe quantum field states in one spatial dimension \cite{Ve10}, 
and they were shown to be a natural continuous version of MPS. 
In this work,``tensor network ansatz'' is synonymous with ``matrix product ansatz'', which means that the coefficients of the state can be expressed as a product of matrices in some basis. 
Further studies about cMPS can be found in Refs.~\cite{Ha13,Je15,Ga18,Os19,Ti19}.

Recently this topic was studied from the opposite perspective \cite{De17b}. 
Namely, one asked the question: 
given a translationally invariant MPS, does it have a continuum limit?
And, if it does, what does the state at the continuum look like?
These questions obviously depend on how one defines a continuum limit---a very subtle question without an unambiguous answer. 
One could argue, though, that the choice made in Ref.\ \cite{De17b} is fairly natural: 
it was defined as the limit of repeatedly applying the inverse of the renormalisation operation considered in Ref.\ \cite{Ve05}, 
together with a regularisation condition in the limit. 
(Henceforth, when we refer to continuum limit of an MPS we mean the one defined in \cite{De17b}.) 
Ref.\ \cite{De17b} fully characterised the set of MPS with a continuum limit, 
and found that the state at the continuum can generally not be expressed as a cMPS. In other words, the class of cMPS was too narrow to express all continuum limits of MPS.

To be more explicit, Ref.\ \cite{De17b} showed that a translationally invariant MPS has a continuum limit if and only if its transfer matrix $E$ is an infinitely divisible quantum channel. 
The latter are channels of the form $E=P e^L$, where $P^2=P$ is a projector quantum channel, $L$ is a Liouvillian of Lindblad form, and $PL=PLP$. 
Since the transfer matrix of a (homogeneous) cMPS is given by $e^L$, 
this class can only express the continuum limits of MPS with a trivial projector, i.e.\ $P=I$, the identity matrix. 
The new element that needs to be represented, thus, is the projector $P$. 
Note that a non-trivial $P$ implies that the transfer matrix has eigenvalues 0 (in particular, the presence of $P$ is unrelated to the degeneracy of the eigenvalue 1, which is related to the (lack of) injectivity of an MPS \cite{Pe07,Sc10b}). 
A simple example of an MPS with a continuum limit 
is the superposition of ferromagnetic states $|0\ldots 0\ra +|1\ldots 1\ra$; its transfer matrix is $E=P=|00\ra\la 00|+|11\ra\la 11|$, and can thus not be expressed as a cMPS. 
This gives rise to the question: 
Is there a matrix product ansatz that can express the continuum limit of any MPS?

In this work, we present a generalised ansatz of cMPS that is capable of expressing the continuum limit of any MPS (\cref{thm:main}). 
Our ansatz consists of a sum of cMPS, each with a different boundary operator, and attached to an ancilla state. 
The boundary operators are given by the Kraus operators of $P$. 
We show that this new ansatz can be understood as the concatenation of 
a state at the closure of the set of cMPS (which will contribute $P$ to the transfer matrix; \cref{thm:thermo}), 
and a standard cMPS on a finite segment (which will contribute $e^L$ to the transfer matrix). 
The former can be thought of as a cMPS with matrices of unbounded norm, 
or a cMPS on a segment of unbounded length, that is, in the thermodynamic limit. 
Informally, this can be understood as the enforcement of some ``superselection rules'' which correspond to the zeros in the transfer matrix brought in by $P$.
Our ansatz thus provides a way to directly enforce them without the need to resort to cMPS with matrices of very large norm, or to very large system sizes, as explained in \cref{ssec:interp}.

This paper is organized as follows. In \cref{sec:channels}, \cref{sec:mps} and \cref{sec:cmps} we present the preliminary material for this work:
 the preliminaries on quantum channels, 
on Matrix Product States, 
and on continuous Matrix Product States, respectively. 
The core of this work is presented in \cref{sec:ansatz}, 
where we present the ansatz of generalised cMPS, we discuss it, provide an interpretation and some examples. 
Finally, we conclude and present an outlook in \cref{sec:conclusions}.
\cref{app:thermo} contains the proof of \cref{thm:thermo},
and \cref{app:PLP} a characterisation of the condition $PL=PLP$ for a special case. 

\section{Preliminaries on quantum channels}
\label{sec:channels}

In this section we present the background on quantum channels. 
We will first present the basic definitions (\cref{ssec:def}), 
then characterise Markovian channels (\cref{ssec:markovian}),
projector quantum channels (\cref{ssec:projector}), 
and finally infinitely divisible channels (\cref{ssec:infdiv}).

\subsection{Basic definitions}
\label{ssec:def}

Throughout this paper we let $\mc{M}_{D,D'}$  denote the set of complex matrices of size $D\times D'$, and $ \mc{M}_D:=\mc{M}_{D,D} $.
 $A\geq 0$ denotes that $A$ is positive semidefinite, i.e.\ Hermitian and with non-negative eigenvalues, and $A>0$ that it is positive definite, i.e., Hermitian with positive eigenvalues.

A quantum channel $\E:\mc{M}_D\to \mc{M}_D$ is a linear, completely positive, trace-preserving map. It has a Kraus decomposition $\E(X)= \sum_{i=1}^d A^i X A^{i \dagger} $, where $^{\dagger}$ denotes complex conjugate transposed. 
The minimal number of Kraus operators $\{A^i\}_{i}$  in this decomposition is called the Kraus rank of $\E$. 

The superoperator $\E$ can be represented as a matrix $E$ by identifying $\mc{M}_D$ with $\mathbb{C}^{D^2}$. Specifically, a matrix $X=\sum_{i,j}X_{i,j}|i\ra\la j|$ is expressed as $|X\ra = \sum_{i,j}X_{i,j}|i\ra |j\ra$, so that the channel $\mc{E}$ becomes $E=\sum_{i=1}^d A^i\otimes \bar A^i$, where $\bar{\;}$ denotes complex conjugate. 
In this representation, composition of channels $\mc{E}_1\circ \mc{E}_2$ becomes the product of the corresponding matrices, $E_1E_2$.
We will often switch from one to the other, always denoting the superoperator version with calligraphic fonts and the operator version with roman fonts.

Finally, we will denote the identity matrix of size $n\times n$ by $I_n$ (or simply $I$, when clear from the context), and the Pauli matrices by $\sigma_x = |0\ra\la 1|+|1\ra\la 0|$, $\sigma_z= |0\ra\la 0|-|1\ra\la 1|$ 
and $\sigma_y=-i|0\ra\la 1| +i |1\ra\la0|$. 

\subsection{Markovian channels}
\label{ssec:markovian}

A quantum channel $\mc{E}$ is called \emph{Markovian} (sometimes also called a \emph{quantum dynamical semigroup}) if $\mc{E} = e^{\mc{L}}$ where $\mc{L}$ is a Liouvillian of Lindblad form \cite{Li76}. 
There are several equivalent representations of $\mc{L}$ \cite{Wo11}, and here we present one of them. 

\begin{definition}[Liouvillian]\label{def:L}
A Liouvillian of Lindblad form $\mc{L}:\mathcal{M}_D\to\mathcal{M}_D $ is a superoperator of the form
\be
\mc{L}[Q,\{R_\alpha\}]= Q \rho +\rho Q^\dagger + \sum_{\alpha=1}^qR_{\alpha} \rho R_{\alpha}^\dagger,
\label{eq:L}
\ee
where 
\be
Q=-iH -\frac{1}{2}\sum_{\alpha=1}^q R_{\alpha}^\dagger R_{\alpha}, 
\label{eq:Q}
\ee
and $H$ is a Hermitian operator. 
\end{definition}

In this context, $H$ is called the Hamiltonian, and $R_{\alpha}$ the jump operators of $\mc{L}$.

\subsection{Projector quantum channels}
\label{ssec:projector}

We say that $\mc{P}:\mc{M}_D\to \mc{M}_D$ is a \emph{projector quantum channel} if it is a quantum channel and it fulfills $\mc{P}(\mc{P}(\rho)) = \mc{P}(\rho)$ for all $\rho \in \mc{M}_D$. Examples of projector quantum channels are:
\begin{enumerate}
\item[(i)] The identity channel, $\mathrm{id}(\rho) = \rho$, 
\item[(ii)] The `pinching map', $\E(\rho) = \sum_{i=1}^D |v_i\ra \la v_i| \rho |v_i\ra \la v_i|$, where $\{|v_i\ra\}$ is some orthonormal basis, and
\item[(iii)] The completely depolarising map, $\E(\rho) = \mathrm{tr}(\rho) \sigma$, where $\sigma> 0$ with $\mathrm{tr}(\sigma)=1$. 
\end{enumerate}
In fact, these three are the building blocks of any projector quantum channel, as we will see below.
To this end, let us first review a characterisation of the fixed point set of a quantum channel. 

\begin{theorem}[Theorem 6.14 of \cite{Wo11}]\label{thm:Pcharac1}
Let $\E:\mc{M}_D\to \mc{M}_D$ be a completely positive, trace-preserving, linear map. Then there is a unitary $U \in \mc{M}_D$ and a set of positive definite density matrices $\sigma_k\in \mc{M}_{m_k}$ such that the fixed point set of $\E$, $\mc{F}_\E:=\{X\in\mc{M}_D\: |\: \E(X)=X\}$, is given by
\be
\mc{F}_\E = U(0\oplus \bigoplus_{k=1}^n\mc{M}_{D_k} \otimes \sigma_k)U^\dagger, 
\label{eq:FE}
\ee
for an appropriate basis of the Hilbert space $\mathbb{C}_D = \mathbb{C}_{D_0}\oplus \bigoplus_{k=1}^n \mathbb{C}_{D_k} \otimes \mathbb{C}_{m_k}$. 
\end{theorem}

We now characterise the action of a projector quantum channel  $\mc{P}$ on a general element $\rho\in \mc{M}_D$. 
Let $\mc{P}:\mc{M}_D\to\mc{M}_D $ be a projector quantum channel whose fixed point set is given by \eqref{eq:FE}. 
Since it is a projector, its fixed point set equals its image. 
Define orthogonal projectors $\pi_k\in\mc{M}_D$ onto the $k$th block in the image of $\mc{P}$, so that 
\be
\mc{P}(\rho) = \sum_{k=1}^n \pi_k \mc{P}(\rho) \pi_k 
=:\bigoplus_{k=1}^n \mc{P}_k(\rho_k).
\ee 
Note that $\pi_0$ has support on a subspace of dimension ${D_0}$, 
and  $\pi_k$ (for $k>0$) has support on a subspace of dimension ${D_k m_k}$.  
We have that 
\be
\mc{P}_k:\mc{M}_{D_k} \otimes \mc{M}_{m_k} \to \mc{M}_{D_k} \otimes \mc{M}_{m_k}
\ee
is a completely positive trace preserving map, and $\rho_k\in  \mc{M}_{D_k}\otimes  \mc{M}_{m_k}$. 
Because of the latter, $\rho_k$ can be decomposed as a sum of elementary tensor factors, i.e.\ 
\be
\label{eq:rho2}
\rho_k = \sum_{l=1}^{r_k} \rho_{k,l}^{(1)}\otimes \rho_{k,l}^{(2)},
\ee
where $\rho_{k,l}^{(1)}\in \mc{M}_{D_k}$ and $\rho_{k,l}^{(2)}\in\mc{M}_{m_k}$, 
and where $r_k$ is the minimal number of terms (called the \emph{operator Schmidt rank} in, e.g., \cite{De19}). 
While decomposition \eqref{eq:rho2} is non-unique, the minimal number of terms is unique (see again, e.g., \cite{De19}). 
Using the three building blocks mentioned at the beginning of this subsection, and the linearity of $\mc{P}_k$, 
it is immediate to see that 
\be
\mc{P}_k (\rho_k) = \sum_{l=1}^{r_k} 
\mathrm{id}(\rho_{k,l}^{(1)})\otimes \mathrm{tr}(\rho_{k,l}^{(2)}) \sigma_{k}.
\ee

In summary, a projector quantum channel $\mc{P}$ whose fixed point set is given by \eqref{eq:FE} acts on an arbitrary element $\rho\in \mc{M}_D$ as follows. 
Let  $V_k\in \mc{M}_{D,D_km_k}$ be the isometry 
 which is the identity map on the support of $\pi_k$, 
 i.e.\ $V^\dagger_kV_k=I_{D_{k}m_k}$ and 
 $V_k V_k^\dagger \pi_k = \pi_k$.
For an element 
\begin{subequations}
\be
\rho &=& \sum_{k,k'=0}^n \pi_k \rho\pi_{k'}\\
 \rho_k   &:=&  V_k^\dagger \pi_k \rho\pi_{k} V_k = \sum_{l=1}^{r_k} \rho_{k,l}^{(1)}\otimes \rho_{k,l}^{(2)}  
\ee
\end{subequations}
we have that 
\be
\mc{P}(\rho) =
\sum_{k=1}^n \pi_k   
V_k\left[
\sum_{l=1}^{r_k} 
\mathrm{id}(\rho_{k,l}^{(1)})\otimes \mathrm{tr}(\rho_{k,l}^{(2)}) \sigma_{k} \right] V_k^\dagger
\pi_k ,
\label{eq:P1}
\ee
where $\sigma_k>0$ and $\mathrm{tr}(\sigma_k)=1$. 


Note that the identity map corresponds to $n=1$, $D_{1}=D$ and $m_{1}=1$;
the pinching map corresponds to $n = D$ with $D_{k} = m_{k} = 1$ for all $k$; 
and the completely depolarising map corresponds to $n = 1$, $D_{1} = 1$ and $m_1= D$.

\subsection{Infinitely divisible quantum channels}
\label{ssec:infdiv}

Having defined Markovian channels and projector quantum channels, we are now ready to define infinitely divisible channels. 
A quantum channel $\mc{E}$ is called \emph{infinitely divisible} \cite{Wo08} (see also Refs.~\cite{Ho87,De89b}) if for every natural $n$ there exists a quantum channel $\mc{E}_n$ such that $\E= (\mc{E}_n)^n$, where the power   stands for $n$-fold composition, $\mc{E}_{n}\circ \cdots \circ \mc{E}_{n}$ ($n$ times). 
We have the following characterisation.

\begin{theorem}[Infinitely divisible channels \cite{Ho87,De89b,De17b}]
\label{thm:infdiv} Let $\E$ be a quantum channel. 
The following are equivalent:
\ben
\im $\E$ is infinitely divisible, i.e.\ $\mc{E}= (\mc{E}_n)^n$ for all $n\in \mathbb{N}$.
\im There is an integer $p>1$ such that
 $\mc{E} =(\mc{E}_{p^{\ell}})^{p^\ell} $ for all $\ell \in \mathbb{N}$, and for all sequences $\{n_k,\ell_k\}_{k=1}^{\infty}$ fulfilling $\lim _{k\to\infty}n_k/p^{\ell_k} = 0$, $(\mc{E}_{n_k})^{p^{\ell_k}}\to \mc{P}$, where $\mc{P}$ is a projector quantum channel.
\im $\mc{E} = \mc{P}e^{\mc{L}}$ where $\mc{P}$ is a projector quantum channel and $\mc{L}$ is a Liouvillian of Lindblad form such that $\mc{P} \mc{L} = \mc{P} \mc{L} \mc{P}$.
\een
\end{theorem}
Therefore, Markovian quantum channels are infinitely divisible channels with $\mc{P}=\textrm{id}$, the identity channel.

\section{Preliminaries on matrix product states}
\label{sec:mps}

In this section we present the background on Matrix Product States (MPS) and their continuum limits required for this paper. 
We will first present the basic definitions of MPS (\cref{ssec:mps}) 
and then review the results on their continuum limits (\cref{ssec:cl}). 

\subsection{Matrix Product States}
\label{ssec:mps}

In this work we consider exclusively translationally invariant MPS. 
Namely, we are given a tensor $A=\{A_i^{\alpha,\beta}\}$, where $A_i^{\alpha,\beta}$ is a complex number. 
The index $i=1,\ldots, d$ 
(or $i=0,1,\ldots, d-1$, depending on the example) is called the physical index and $d$ the physical dimension.
The indices $\alpha,\beta=1,\ldots,D$ 
(or $\alpha,\beta=0,1,\ldots, D-1$, depending on the example) 
are called virtual indices and $D$ the bond dimension. 
This tensor defines a translationally invariant MPS \footnote{This is in fact a Matrix Product Vector \cite{Ci17}, but in this paper we will ignore this distinction and denote unnormalized Matrix Product Vectors as MPS.}
\be
|V_N(A)\ra = \sum_{i_1,\ldots, i_N=1}^d \mathrm{tr}(A_{i_1}A_{i_2}\cdots A_{i_N})|i_1,i_2,\ldots i_N\ra,
\label{eq:VNA}
\ee
and a family of MPS
\be
\mc{V}(A) = \{|V_N(A)\ra\}_N.
\ee

$|V_N(A)\ra$ describes the state of, for example, a spin chain of $N$ spins, each of which is described by a $d$-dimensional vector space. Among the spins there is a fixed lattice spacing $a$, which is arbitrary but fixed, so that $|V_N(A)\ra$ is defined on a segment of length $l_N:=Na$. 
Note, thus, that what is common in the family $\mc{V}(A)$ is the lattice spacing $a$. 

The \emph{transfer matrix} associated to $\mc{V}(A)$ is given by $E_a=\sum_{i=1}^d A_i\otimes \bar A_i$.
The subindex $a$ emphasises the length over which it is defined. 
$E_a$ is the matrix representation of the completely positive map $\E_a(\rho) =\sum_{i=1}^d A_i \rho A_{i}^\dagger$, and one can assume without loss of generality that this is trace preserving \cite{De17b,De17}. 
Hence we will refer to $E_a$ as a quantum channel.

\subsection{The continuum limit of an MPS}
\label{ssec:cl}

We review here some definitions and results of Ref.~\cite{De17b}. 
First, a family of MPS $\mc{V}(A)$ can be $p$-\emph{refined} if there a tensor $B$ and an isometry $W:\mathbb{C}^d\to (\mathbb{C}^d)^{\otimes p}$ such that 
\be
W^{\otimes N}|V_N(A)\ra =|V_{pN}(B)\ra\quad \forall N.
\ee
Thus, one $p$-refinement step divides the lattice spacing by $p$, 
$a\mapsto a/p$, and multiplies the number of sites by $p$, namely
$N\mapsto Np$, so that $l_N = aN$ remains unchanged. 
This happens for all $N$ in parallel.

As argued in Ref.~\cite{De17b}, it is not satisfactory to define the continuum limit as the infinite iteration of the $p$-refining procedure, but one additionally needs to impose that the limit be stable under the blocking of a few spins.

\begin{definition}[Continuum limit of an MPS \cite{De17b}]
\label{def:cl}
A family of MPS $\mc{V}(A)$ has a continuum limit if there is a $p>1$ such that $\mc{V}(A)$ can be $p$-refined infinitely many times, and blocking $n_k$ spins after $\ell_k$ $p$-refining steps yields the same result, as long as $(n_k/p^{\ell_k})\to 0$. 
\end{definition}

The set of MPS with a continuum limit is fully characterised as follows. 

\begin{theorem}[Continuum limit of an MPS \cite{De17b}]
\label{thm:cl}
$\mc{V}(A)$ has a continuum limit if and only if its transfer matrix $E_a$ is infinitely divisible. 
\end{theorem}

Using the characterisation of \cref{thm:infdiv}, and choosing the appropriate normalisation of $L$, we thus have that $\mc{V}(A)$ has a continuum limit if and only if $E_a = Pe^{aL}$, where $P^2=P$, and $PL=PLP$. 

\section{Preliminaries on continuous Matrix Product States}
\label{sec:cmps}

In this section we present the background on cMPS, mainly by following Ref.~\cite{Ha13}. 
We will first present the mathematical setting (\cref{ssec:tmcmps}) and 
then the definition of continuous MPS (\cref{ssec:cmps}).

\subsection{Mathematical setting}
\label{ssec:tmcmps}

We first define a segment $\mc{R}$ as an interval of the reals $\mc{R}=[x,y]$, where $x<y$ and $x,y\in \mathbb{R}$. We will consider a quantum system defined on $\mc{R}$, which accommodates $q$ bosonic or fermionic particle species, which are labeled by the index $\alpha=1,\ldots,q$. 
The $N$ fold cartesian product of $\mc{R}$ is denoted $\mc{R}^N$. 
The symmetric (antisymmetric) subspace of $\mc{R}^N$ is denoted $\mc{R}_{+}^N$ ($\mc{R}_{-}^N$). 
If particle of type $\alpha$ is bosonic (fermionic), then a state of $N_{\alpha}$ particles of type $\alpha$ is described by a square integrable function on $\mc{R}_{\eta_{\alpha}}^{(N_{\alpha})}$, where $\eta_{\alpha}=+1(-1)$, denoted 
$ L^2(\mc{R}_{\eta_{\alpha}}^{(N_{\alpha})})$. 
Thus a state with $N_\alpha$ particles of type $\alpha$, for $\alpha=1,\ldots, q$, is an element of 
\be
\mathbb{H}_{\mc{R}}^{(N_1,\ldots, N_q)} = L^2\Bigl(\prod_{\alpha=1}^q \mc{R}_{\eta_{\alpha}}^{(N_{\alpha})}\Bigr) .
\ee
An arbitrary state of the system is an element of the Fock space
\be
\mathbb{H}_{\mc{R}} = 
\bigoplus_{N_1=0}^\infty \cdots 
\bigoplus_{N_q=0}^\infty
\mathbb{H}_{\mc{R}}^{(N_1,\ldots, N_q)}. 
\ee
We refer to this space as the physical space.
$|\Omega_{\mc{R}}\ra$ denotes the vacuum state, i.e. $|\Omega_{\mc{R}}\ra\in \mathbb{H}_{\mc{R}}^{\{N_\alpha=0\}} $. 

Now, a particle of type $\alpha$ is created or annihilated at position $x\in \mc{R}$ with the operators $\hat\psi^\dagger_{\alpha}(x)$ and $\hat\psi_{\alpha}(x)$, respectively. These satisfy the commutation or anticommutation relations
\be
&&\hat\psi_{\alpha}(x)\hat\psi_{\beta}(y)-\eta_{\alpha,\beta} \hat\psi_{\beta}(y)\hat\psi_{\alpha}(x)=0, \\
&&\hat\psi_{\alpha}(x)\hat\psi^\dagger_{\beta}(y)-\eta_{\alpha,\beta} \hat\psi^\dagger_{\beta}(y)\hat\psi_{\alpha}(x)=\delta_{\alpha,\beta}\delta(x-y),
\ee
where $\eta_{\alpha,\beta}=-1$ if both $\alpha$ and $\beta$ represent fermionic particles, and $\eta_{\alpha,\beta}=1$ when at least one of the two particles is bosonic; clearly, $\eta_{\alpha,\alpha}=\eta_{\alpha}$.

The auxiliary space is $\mathbb{C}^D$, where $D$ is the bond dimension. The variational parameters of the cMPS (\cref{def:cmps}) will correspond to the functions $Q,R_{\alpha}: \mc{R} \to \mathcal{B}(\mc{M}_D)$,
 that take value in $ \mathcal{B}(\mc{M}_D)$, the space of bounded linear operators acting on the auxiliary space. 
In addition, the boundary operator $B \in \mathcal{B}(\mc{M}_D) $ will encode the boundary conditions.

\subsection{Definition of continuous MPS}
\label{ssec:cmps}

In this work we focus exclusively on homogeneous cMPS \cite{Ha13}, and we  refer to them simply as cMPS. In this case, the matrices $Q$ and $\{R_\alpha\}$ do not depend on the position $x$. 

\begin{definition}[cMPS]\label{def:cmps}
A cMPS on a segment $\mc{R}$ is defined as 
\be
&&|\phi_\mc{R}[B, Q,\{R_{\alpha}\}]\ra = \nonumber\\
&&\mathrm{tr}_{\mathrm{aux}}\Big\{B\: 
\mc{T}\!\exp \Big[\int_\mc{R}dx \Big(Q\otimes I + \sum_{\alpha=1}^q R_{\alpha}\otimes \hat\psi_\alpha^\dagger (x)\Big)\Big]\Big\}|\Omega_{\mc{R}}\ra,\nn\\
\ee
where $B,Q,R_{\alpha}\in \mc{B}(\mc{M}_D)$ and $\mc{T}\!\exp$ denotes the path ordered exponential. 
\end{definition}
Note that $|\phi_\mc{R}[B, Q,\{R_{\alpha}\}]\ra \in \mathbb{H}_{\mc{R}}$, which is the physical space. 
We also define the corresponding operator which lives in the auxiliary and physical space 
$\phi_\mc{R}[B,Q,\{R_{\alpha}\}] \in \mc{M}_D\otimes \mathbb{H}_{\mc{R}}$ as 
\be
&&\phi_\mc{R}[B,Q,\{R_{\alpha}\}] =\nn\\
&&=(B\otimes I)\: \mc{T}\!\exp 
\Big[\int_\mc{R} dx \Big(Q\otimes I + \sum_{\alpha=1}^q R_{\alpha}\otimes \hat\psi_\alpha^\dagger (x)\Big)\Big]\Big\}|\Omega_\mc{R}\ra, \quad \nn\\
\label{eq:cmpsopen}
\ee
so that 
\be
|\phi_{\mc{R}}[B,Q,\{R_{\alpha}\}]\ra =\mathrm{tr}_{\textrm{aux}}(\phi_{\mc{R}}[B,Q,\{R_{\alpha}\}]).
\ee
We will say that $\phi_\mc{R}[B,Q,\{R_{\alpha}\}]$ has ``open auxiliary indices." 
We also define an inner product for these objects
\be
(\cdot,\cdot): 
(\mc{M}_D \otimes \mathbb{H}_{\mc{R}}, \mc{M}_D \otimes \mathbb{H}_{\mc{R}}) \to 
\mc{M}_D \otimes \mc{M}_D,
\label{eq:scalarprod}
\ee
as 
\be
&&(\phi_\mc{R}[B',Q',\{R'_\alpha\}],\phi_\mc{R}[B, Q,\{R_\alpha\}]) = \nonumber\\
&&(B'\otimes \bar B) \: \exp \left[ |\mc{R}|
\left( Q'\otimes I +I\otimes \bar Q + \sum_{\alpha=1}^q R'_{\alpha}\otimes \bar R_{\alpha}\right)\right]. 
\ee
With this, and by analogy with the discrete case, we  define the transfer matrix of the state $|\phi_\mc{R}\ra$ for a length $|\mc{R}|=a$ as 
\be
E_a &=& (\phi_\mc{R}[B, Q,\{R_\alpha\}],\phi_\mc{R}[B, Q,\{R_\alpha\}]) \nn\\
&=& (B\otimes \bar B) e^{ a L[Q,\{R_\alpha\}]},
\label{eq:tm}
\ee
where $L[Q,\{R_\alpha\}]$ is the matrix version of the Liouvillian of Lindblad form of Eq.~\eqref{eq:L}. 
Note that the transfer matrix of a cMPS (with boundary conditions $B=I$) is a Markovian quantum channel, where the jump operators of the Liouvillian are precisely $\{R_{\alpha}\}$, and the Hamiltonian is determined by $Q$ via \eqref{eq:Q}.

\section{Representing the continuum limit of an MPS}
\label{sec:ansatz}

In this section we present the core of this work, namely a generalised ansatz of cMPS. 
First we will precisely state the problem (\cref{ssec:statement}),
then we will present a generalised ansatz of cMPS addressing this problem (\cref{ssec:gencmps}),
we will discuss it (\cref{ssec:discussion}),
provide an interpretation thereof (\cref{ssec:interp})
and finally give some examples (\cref{ssec:ex}).

\subsection{Statement of the problem}
\label{ssec:statement}

We saw in \cref{thm:cl} that a family of MPS $\mc{V}(A)$ has a continuum limit if and only if its transfer matrix is infinitely divisible, i.e.\ of the form $E_a=Pe^{aL}$, where $P^2=P$ and $PL=PLP$. 
On the other hand, 
we saw in Eq.~\eqref{eq:tm} that the transfer matrix of a cMPS is Markovian, i.e.\ of the form $e^L$.  
Thus, cMPS can only represent the continuum limit of MPS whose transfer matrix is Markovian. 
An example of this lack of generality is the equal superposition of two ferromagnetic states $\ket{0,0,\dots,0}+\ket{1,1,\dots,1}$, whose transfer matrix is $E_a=P=|0,0\ra\la 0,0| + |1,1\ra \la 1,1|$, and thus its continuum limit cannot be expressed as a cMPS.
A generalised cMPS (to be presented next) is a matrix product ansatz, defined directly in the continuum, which is capable of expressing the continuum limit of any MPS.

\subsection{A generalised ansatz of cMPS}
\label{ssec:gencmps}

We now present the central result of this work. 

\begin{theorem}[Main result]\label{thm:main}
Let $\mc{V}(A)=\{|V_N(A)\ra\}_N$ be a family of MPS which has a continuum limit according to \cref{def:cl}, and let its transfer matrix be $E_a = Pe^{aL[Q,\{R_\alpha\}]}$, 
where $P^2=P$ is a projector quantum channel, 
$L[Q,\{R_\alpha\}]$ a Liouvillian of Lindblad form,
 and $PL=PLP$. 
Consider a Kraus decomposition of $P$
\be
\label{eq:P}
P = \sum_{i=1}^K B_i \otimes \bar B_i, 
\ee
where $K$ is the Kraus rank of $P$.
Then, for any $N$, the continuum limit state of $|V_N(A)\ra$ can be represented by the \emph{generalised cMPS} 
\be
\label{eq:generalisedcmps}
\ket{\Phi_{\mc{R}}[\{v_i\},\{B_i\},Q,\{R_\alpha\}]}=\sum_{i=1}^K \ket{v_i}\otimes\ket{\phi_{\mc{R}}[B_i,Q,\{R_\alpha\}]},
\ee
where $\{\ket{v_i}\}_{i=1}^K$ is an orthonormal basis of $\mathbb{C}^K$, and $\ket{\phi_{\mc{R}}[B_i,Q,\{R_\alpha\}]}\in\mathbb{H}_{\mc{R}}$ is a cMPS (\cref{def:cmps}). 
\end{theorem}

If clear from the context, we shall simply write $|\Phi_{\mc{R}}\ra$ for the generalised cMPS, or 
$\Phi_\mc{R}$  for the corresponding generalised cMPS with open auxiliary indices.  

Note that $\ket{\Phi_{\mc{R}}} \in \mathbb{C}^K \otimes \mathbb{H}_{\mc{R}}$. 
In words, the generalised cMPS is a sum of cMPS, all with the same $Q$ and $\{R_\alpha\}$, but with boundary matrices $B_i$ given by the Kraus operators of $P$. 
In addition, each such cMPS is attached to a state $|v_i\ra \in\mathbb{C}^K $, the \emph{ancilla} space. 
(Note that $\mathbb{C}^D$ is called the \emph{auxiliary} space, and this is already present in a cMPS, cf.\ \cref{ssec:cmps}).
The ancilla space is the novelty in comparison with cMPS. 
Indeed, for $P=I$ (the identity channel), we have $K=1$ and thus recover the case of a cMPS with periodic boundary conditions ($B=I$). 

Note also that $\ket{\Phi_{\mc{R}}}$ is a matrix product, as can be seen by expanding the time ordered exponential as a sum of states in $\mathbb{H}_{\mc{R}}$, each of which has a finite number of excitations with respect to the vacuum.

\begin{figure}[h]\centering
\includegraphics[width=\columnwidth]{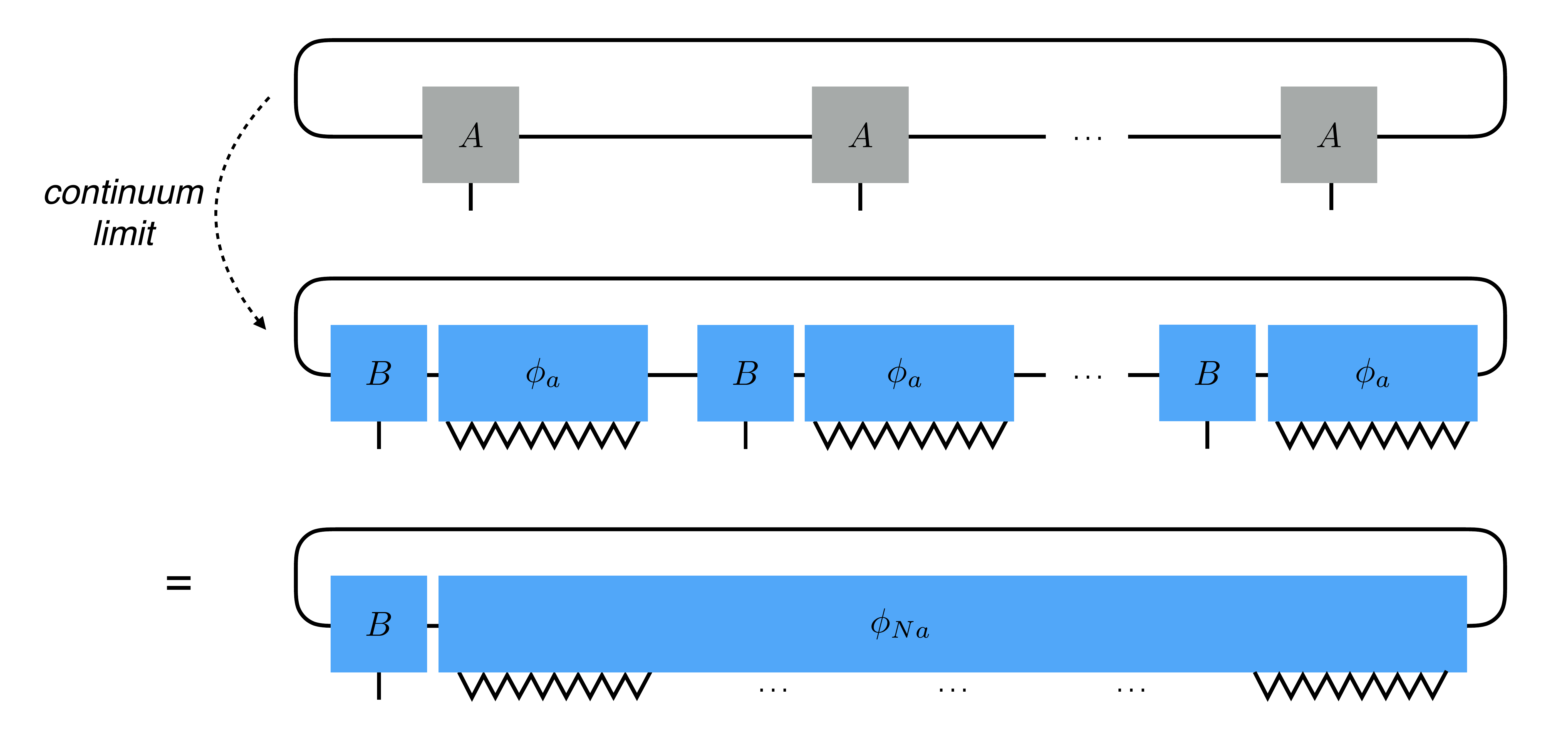}
\caption{
(Top) 
Graphical representation of the MPS $|V_N(A)\ra$ [Eq.\ \eqref{eq:VNA}], which by assumption has a continuum limit, i.e.\ its transfer matrix is $E_a = Pe^{aL}$. 
(Middle) 
Graphical representation of the continuum limit of  $|V_N(A)\ra$, where  
$\{B_i\}$ are the Kraus operators of $P$ [Eq.\ \eqref{eq:P}], 
and $\phi_a$ is shorthand   for $\phi_{[ra,(r+1)a]}[I,Q,\{R_{\alpha}\}]$ [Eq.\ \eqref{eq:cmpsopen}], 
where $r=0,\dots, N-1$, and the continuous index is represented as serrated. 
(Bottom)  Using the fact that $PL=PLP$ and $P^2=P$, the previous line is equivalent to this expression. 
}
\label{fig:MPSlimit}
\end{figure}

Finally, note that conditions $P^2=P$ and $PL=PLP$ guarantee that $(E_a)^N = P e^{aNL} =E_{Na}$, and thus any position $x\in[0,Na)$ is well defined (see \cref{fig:MPSlimit}).

\begin{proof}[of \cref{thm:main}]
To compute the transfer matrix over a segment $\mc{R}$ of length $|\mc{R}|=a$, 
we define $\Phi_{\mc{R}}$ with open auxiliary indices in analogy with Eq.\ \eqref{eq:cmpsopen}
 as
\be
\Phi_{\mc{R}}[\{v_i\},\{B_i\},Q,\{R_\alpha\}]=\sum_{i=1}^K \ket{v_i}\otimes\phi_{\mc{R}}[B_i,Q,\{R_\alpha\}], 
\ee
where $\phi_{\mc{R}}[B_i,Q,\{R_\alpha\}] \in  \mc{M}_D \otimes\mathbb{H}_{\mc{R}}$ is a cMPS with open auxiliary indices, and   $\Phi_{\mc{R}}\in\mathbb{C}^K\otimes\mc{M}_D\otimes\mathbb{H}_{\mc{R}}$. 
We define the scalar product, which we denote $(\cdot,\cdot)$ (not to be confused with the scalar product of \eqref{eq:scalarprod}; see following lines) as
\be
(\mathbb{C}^K\otimes\mc{M}_D\otimes\mathbb{H}_{\mc{R}}) \times (\mathbb{C}^K\otimes\mc{M}_D\otimes\mathbb{H}_{\mc{R}}) \to\mc{M}_D\otimes\mc{M}_D
\ee
as the usual scalar product in quantum mechanics $\mathbb{C}^K \times \mathbb{C}^K \to \mathbb{C}$ 
times the scalar product 
$(\mc{M}_D\otimes\mathbb{H}_{\mc{R}})\times (\mc{M}_D\otimes\mathbb{H}_{\mc{R}}) \to \mc{M}_D\otimes\mc{M}_D$ defined in \eqref{eq:scalarprod}.  
We thus obtain that
\be
E_a &=& (\Phi_{\mc{R}},\Phi_{\mc{R}}) \nn\\
&=& \sum_{i,j=1}^K \la v_j |v_i\ra\left(B_j\otimes \bar B_i\right) \cdot \left(\phi_{\mc{R}}(B_j,Q,\{R_\alpha\}),\phi_{\mc{R}}(B_i,Q,\{R_\alpha\})\right)\nn\\
&=&\left( \sum_{i=1}^KB_i\otimes \bar B_i\right) e^{a L[Q,\{R_\alpha\}] },
\ee
where we have used that $\{\ket{v_i}\}$ is an orthonormal basis. 
Using \eqref{eq:P} we obtain that the transfer matrix is given by $E_a = P e^{a L[Q,\{R_\alpha\}] }$, as we wanted to show. 
\end{proof}

\subsection{Discussion}
\label{ssec:discussion}

We now discuss the ansatz of generalised cMPS presented above.
First, a central observation is that in our setting the transfer matrix is more important than the state at the continuum.
For this reason, it is natural that for a family of MPS $\mc{V}(A)$ with transfer matrix $E_a = Pe^{aL}$ there be many states in the continuum $\ket{\Phi_\mc{R}}$ whose transfer matrix is $(\Phi_{\mc{R}},\Phi_{\mc{R}}) = E_{|\mc{R}|}$. In the following we characterise this freedom. 

First, concerning the choice of $P$ and $L$ we have that:
\begin{enumerate}[label=(\alph*)]
\item Given $E_a$, the projector $P$ is fixed, because it determines the kernel of $E_a$, as $e^L$ has no kernel.
\item If there is Liouvillian of Lindblad form $L$ such that $E_a = Pe^{aL}$ and $PL = PLP$, then any other Liouvillian $L'$ such that $PL = PL'$ will satisfy the same conditions.
\end{enumerate}

Second, once $P$ and $L$ are fixed, there is the following freedom:
\begin{enumerate}[label=(\alph*)]
\item Given $P$, there is the freedom of the choice of Kraus operators of $P$. 
Since we only admit decompositions with the minimal number of terms $K$, 
all such Kraus operators are related by a unitary operation acting on the physical index, i.e.\ $\tilde{B}_j =\sum_{j=1}^K U_{j,i} B_i$. 
\item Given $L$, there is the freedom of the choice of $Q$ and $\{R_{\alpha}\}_{\alpha=1}^q$, characterised in \cite[Proposition 7.4]{Wo11}. 
Here we only consider decompositions where the number of jump operators $q$ is minimal, as this corresponds to the number of particle species of the cMPS.
\end{enumerate}

In addition, once these features of $P$ and $L$ are fixed, there is the following freedom in choosing the space where $|\Phi_\mc{R}\ra $ lives:

\begin{enumerate}[label=(\alph*)]
\item The ancilla space associated to $P$ is $\mathbb{C}^K$, which is uniquely defined.
However, the basis choice $\{\ket{v_i}\}_i$ is arbitrary. 
\item The physical space associated to $e^{aL}$ is the Fock space $\mathbb{H}_\mc{R}$. This is a not uniquely defined---one can essentially choose any Fock space, i.e.\ one has to choose a vacuum and excitations on top of this vacuum. The number of different excitations $q$ is uniquely fixed, as this is the minimal number of jump operators of $L$.
\end{enumerate}

In summary, 
the continuum limit of a family of MPS $\mc{V}(A)$ with transfer matrix $E_a = Pe^{aL}$ can be written as a generalised cMPS, for which we have to choose:
\begin{enumerate}
\item A Liouvillian of Lindblad form $L$ such that $PL=PLP$ and such that $E_a=Pe^{aL}$.
\item The Kraus operators $\{B_i\}_{i=1}^K$ of $P$. 
\item $Q$ [satisfying \eqref{eq:Q}] and $\{R\}_{\alpha=1}^q$ such that $L=Q\otimes I+I\otimes \bar Q+\sum_{\alpha=1}^q R_{\alpha}\otimes \bar R_{\alpha}$.
\item An orthonormal basis $\{|v_i\ra\}_{i=1}^K$ of $\mathbb{C}^K$. 
\item A Fock space $\mathbb{H}_\mc{R}$.
\end{enumerate}

A second comment concerns \cref{def:cl}, i.e.\ the definition of continuum limit. 
As mentioned above, the equal superposition of ferromagnetic states $\ket{0,0,\dots,0}+\ket{1,1,\dots,1}$ has a continuum limit. 
However, this state is distinguishable from a probabilistic mixture of $|0\ldots 0\ra\la 0\ldots 0|$ and $|1\ldots 1\ra\la 1\ldots 1|$ only if one has access to completely non-local (i.e.\ global) observables, whereas in defining and analysing continuum limits one usually assumes that only the algebra of quasi-local observables is accessible. 
In \cref{def:cl}, 
we ask that there be an infinite sequence of isometries $W_1, W_2,\ldots$ that $p$-refine $|V_N(A)\ra$ (and the same sequence of isometries can be applied to all $N$ uniformly; and additionally a regularisation condition holds in the limit).
This implies that the continuum limit state of $|V_N(A)\ra$ has the same information as $|V_N(A)\ra$ itself, as the refining process consists of a sequence of bases changes. 
It follows that no information is  lost in this refining process, in particular, no global information.
For further interpretations of our continuum limit, see \cref{ssec:interp}.

A third comment is that condition $PL=PLP$ is not included in the ansatz of \eqref{eq:generalisedcmps}, since the $\{B_i\}$, $Q$ and $\{R_\alpha\}$ need not obey any condition. 
For this reason, the class of generalised cMPS is broader than the set of continuum limit states of MPS. 
That is, there may be a narrower class of states which can represent the continuum limit of any MPS, and in which $\{B_i\}$, $Q$ and $\{R_\alpha\}$ need  satisfy $PL=PLP$.

As a fourth comment, let us consider correlation functions with generalised cMPS in a segment $\mc{R} = [0,l]$. 
Using Eq.\ \eqref{eq:P}, and the fact that $Q$ and $\{R_\alpha\}$ are independent of the position $x$, 
we obtain 
\begin{widetext}
\be
G^{\alpha,\beta}(x,y)&=&
\langle\Phi_{\mc R}[\{v_i\},\{ B_i\}, Q, \{  R_{\alpha}\}]\,|\,
\hat\psi_{\alpha}^{\dagger}(x)\,\hat\psi_{\beta}(y)\,|\,
\Phi_{\mc R}[\{v_i\},\{ B_i\}, Q, \{ R_{\alpha}\}]\rangle\nn\\
&=&\theta(x-y)\;
\mathrm{tr}_{\text{aux}}
\left[P
e^{y L_{\alpha,\beta}}
\Big(R_{\beta}(y)\otimes I\Big)
e^{(x-y) L_{\alpha}} 
\Big(I\otimes\bar R_{\alpha}(x)\Big)
e^{(l-x)  L} \right] \nn \\
&+& \theta(y-x) \;
\mathrm{tr}_{\text{aux}}
\left[P
e^{x  L_{\beta,\alpha}}
\Big(I\otimes\bar R_{\alpha}(x)\Big) 
e^{(y-x) L_{\beta}} 
\Big(R_{\beta}(y)\otimes I\Big) 
e^{(l-y) L} \right],
\ee
\end{widetext}
where 
\be
L_{\alpha}&=&Q\otimes I+I\otimes \bar Q+\sum_{\beta}\eta_{\alpha,\beta}\: R_{\beta}\otimes\bar R_{\beta} ,\\
L_{\alpha,\beta}&=&Q\otimes I+I\otimes \bar Q+\sum_{\gamma}\eta_{\alpha,\gamma}\eta_{\beta,\gamma}\: R_{\gamma}\otimes\bar R_{\gamma}.
\ee
Note that $L_{\alpha,\alpha}=L$ [the matrix version of $\mc{L}$ in Eq.\ \eqref{eq:L}] since $\eta_{\alpha,\beta}^2=1$.
We thus see that the only difference with respect to the correlation function of a cMPS \cite{Ha13} is that the boundary operator $B\otimes \bar B$ is replaced by a sum of  boundary operators, $\sum_i B_i\otimes \bar B_i =P$.

Finally, the generalised cMPS inherits properties from the cMPS giving rise to $e^{|\mc{R}|L}$ in the transfer matrix. 
In particular, since 
$\frac{d}{dx} \hat\psi_\alpha(x) |\Phi_{\mc{R}}\ra =  \sum_{i=1}^K |v_i\ra \otimes \frac{d}{dx} \hat\psi_\alpha(x) |\phi_\mc{R}\ra$, 
the requirement that this object has a finite norm implies that the matrices $Q$, $\{R_\alpha\}$ obey the regularity conditions presented in \cite[Sec.\ III]{Ha13}, which guarantee that the kinetic energy is finite. 
(These `regularity conditions' are  unrelated to those of the continuum limit given in \cref{def:cl}).

\begin{example}[Superposition of ferromagnetic states] \label{ex:ferro}
Consider the superposition of two ferromagnetic states $|V_N(A)\ra=\ket{0\dots0}+\ket{1\dots1}$. 
The transfer matrix is 
$E_a=P= |0,0\ra\la 0,0| +|1,1\ra\la 1,1| $, thus $Q=R_\alpha=0$. 
If we choose the Kraus operators of $P$ to be $B_i = |i\ra \la i|$ for $i=0,1$, 
and some basis of the ancilla space, $|v_0\ra,|v_1\ra$ then 
the continuum limit state can be written as a generalised cMPS 
\be
\ket{\Phi_{\mc{R}}} = (\ket{v_0} + \ket{v_1}) \otimes \ket{\Omega_{\mc{R}}}.
\ee 
If we choose the Kraus operators of $P$ to be $B_0=I$ and $B_1=\sigma_z$, 
and some other basis of the ancilla space, $|w_0\ra,|w_1\ra$ 
then the generalised cMPS is given by 
\be
\ket{\Phi_{\mc{R}}} = \ket{w_0} \otimes \ket{\Omega_{\mc{R}}}.
\ee
We thus see that if $L=0$ then not the entire space $\mathbb{C}^K$ may be occupied, but only part of it 
(i.e.\ we effectively have $K=1$). 
In \cref{ex:bracket} we will consider a channel $E_a=Pe^L$ with the same $P$ as but with $L\neq0$, in which the entire space will be occupied [Eq.\ \eqref{eq:contbracket}]. 
\end{example}

\subsection{Interpretation of the ansatz}
\label{ssec:interp}

In this section we show that every projector quantum channel $P$ can be obtained as the projection to the fixed point subspace of some Markovian channel $e^{\tilde{L}}$, that is, that for every $P$ there is an $\tilde{L}$ such that 
$P=\lim_{t\to\infty} e^{t\tilde{L}}$ (\cref{thm:thermo}).
(We denote this Liouvillian $\tilde{L}$ in order to distinguish it from the Liouvillian $L$ of the infinitely divisible channel, $E_a=Pe^{aL}$). 
Note that, while $e^{t\tilde{L}}$ corresponds to the transfer matrix of a cMPS, $\lim_{t\to\infty} e^{t\tilde{L}}$ does not. Thus, the  state corresponding to $\lim_{t\to\infty} e^{t\tilde{L}}$ is at the closure of the set of cMPS, not a cMPS itself. 

This limit  can be interpreted in two different ways, which are mathematically indistinguishable. 
The first interpretation is a divergence of the norm of the Liouvillian. 
That is, we see $t$ as  the norm of $\tilde{L}$, and the limit is interpreted as considering the same Liouvillian $\tilde{L}$ with larger and larger norm. 
(Formally, we consider a sequence $(L_i)_{n\in\N}$ such that $|| L_i|| \to \infty$, and where $L_i/|| L_i|| $ is independent of $i$. 
We then identify $t$ with $|| L_i||$, and $ L_i/|| L_i|| $ with $\tilde{L}$.)  
The divergence of the norm of the Liouvillian is due to the divergence of the norm of $Q$ or $\{R_\alpha\}$. 
In any case, the effect of this diverging norm is to forbid certain states, 
namely the states not in the kernel of $e^{\tilde{L}}$. 
In other contexts, such enforcements are called `superselection rules'---here we only this word loosely, and we mean what is stated in the previous sentence.

The second interpretation is that $\tilde{L}$ is normalised, but the length of the segment where the corresponding cMPS is defined ($|\tilde{\mc{R}}|$) diverges. This thus corresponds to the thermodynamic limit of a cMPS. 

Together with \cref{thm:main}, this means that the generalised cMPS $|\Phi_{\mc{R}}\ra$ 
can be seen as the concatenation 
of an element at the closure of the set of cMPS , and a cMPS. 
Using the second interpretation,
the first element can be seen as the thermodynamic limit of a cMPS in the thermodynamic limit  (which gives rise to $P$), 
and  the second element is a cMPS defined on a region of length $|\mc{R}|$  (which gives rise to $e^{|\mc{R}|L}$).

Let us now state the main result of this section, namely \cref{thm:thermo}.
We will prove the statement for the corresponding superoperator versions  of $P$ and $\tilde{L}$, namely $\mc{P}$ and $\tilde{\mc{L}}$ (see \cref{ssec:def}).  To this end, 
consider the projector quantum channel $\mc{P}$ given by \eqref{eq:P1}. 
Consider a Liouvillian of Lindblad form 
$\tilde{\mc{L}}[Q,\{R_{k,1},R_{k,2}\}_{k=1}^n]$ [see Eq.~\eqref{eq:L}] given by 
 \begin{subequations}
\be
H=0, 
\ee
\be
R_{k,1}= \ket{k}\bra{k} \otimes I_{D_{k}}\otimes 
\sum_{i=1}^{m_{k}-1}
\theta_{k,i}
|v_{k,i}\ra \la v_{k,i+1}|, 
\ee
\be
R_{k,2}= \ket{k}\bra{k} \otimes I_{D_{k}}\otimes 
\sum_{i=1}^{m_{k}-1}
\theta_{k,i+1}|v_{k,i+1}\ra \la v_{k,i}|, 
\ee\label{eq:Cthermo}\end{subequations}
where $ \{\ket{k}\}$ is the computational basis, $\theta_{k,i}>0$, and $\{|v_{k,i}\ra\}$ is an orthonormal basis. 
Recall that $Q$ is determined by $H$ and $\{R_{k,1},R_{k,2}\}$ as specified in \eqref{eq:Q}.

\begin{proposition}\label{thm:thermo}
Let $\mc{P}$ be the projector quantum channel given by \eqref{eq:P1}. 
Then the Liouvillian of Lindblad form $\tilde{\mc{L}}$ given by \eqref{eq:Cthermo} is such that 
\be
\mc{P} =\lim_{t\to \infty}e^{t \tilde{\mc{L}} [Q,\{R_{k,1},R_{k,2}\}_{k=1}^n]}. 
\label{eq:Plim}
\ee 
\end{proposition}

The proof of \cref{thm:thermo} can be found in \cref{app:thermo}. 
We end with two remarks concerning this result. 
First, note that, given $\mc{P}$, $\tilde{\mc{L}}$ in \cref{thm:thermo} is highly not unique. 
Since $\mc{P}$ is a projection to the kernel of $\tilde{\mc{L}}$, $\textrm{Ker}(\tilde{\mc{L}})$, any other $\mc{L}$ with the same kernel will be equally valid. 
Second, note that if $E_a = Pe^{aL}$  and  $P =\lim_{t\to\infty}e^{t\tilde{L}}$,  
then $L$ and $\tilde{L}$ do not necessarily commute. 
This is because $PLP=PL$ implies that $L$ is lower block triangular in the basis given by 
$\textrm{Ker}(\tilde{L}),(\textrm{Ker}(\tilde{L}))^{\perp}$. 
In \cref{app:PLP} we characterise $P$ and $L$ such that $PL=PLP$ for a special case.

\subsection{Examples}
\label{ssec:ex}

Let us now consider two examples  to illustrate  \cref{thm:thermo}, and one example to illustrate both \cref{thm:thermo} and \cref{thm:main}. 

Recall that in our setting  the starting point is the transfer matrix, and the other quantities, such as the continuum limit state, are derived from it.

\medskip

\begin{example}[Superposition of ferromagnets]
We revisit the equal superposition of ferromagnetic states (\cref{ex:ferro}), this time to illustrate \cref{thm:thermo}. 
We consider in this case the equal superposition of $K$ ferromagnetic states, $|V_N(A)\ra = \sum_{m=1}^K (|m\ra)^{\otimes N}$, whose  transfer matrix is given by $E_a = P= \sum_{m=1}^K |m\ra \la m| \otimes |m\ra \la m|$, and thus has  
a continuum limit. 
Following \cref{thm:thermo} 
(and, specifically, to the interpretation of  the limit as the thermodynamic limit, $|\mathcal{R}|\to\infty$; see \cref{ssec:interp}), 
we consider a cMPS with $H=0$ and jump operators $R_m = |m\ra \la m| $ for $m=1,\ldots,K$.  
This gives rise to the following cMPS with open indices (see \cref{ssec:cmps}) on a segment $\mc{R}$ of length $|\mc{R}|=l$, 
\be
\phi_{\mc{R}} = e^{-|\mc{R}|/2} \left\{I\otimes I + \sum_{m=1}^K R_m \otimes \bigg[\bigg( \mc{T}\!\exp\int_{\mc{R}} dx \: \hat\psi^\dagger_m(x) \bigg)  - I\bigg]\right\}|\Omega_{\mc{R}}\ra. \nonumber\\
\ee
This satisfies that 
\be
E_{|\mc{R}|}= (\phi_{\mc{R}},\phi_{\mc{R}}) = e^{-|\mc{R}|} [I\otimes I + \sum_{m=1}^K R_m\otimes R_m (e^{|\mc{R}|}-1)],  
\ee
which verifies that $\lim_{|\mc{R}|\to\infty}E_{|\mc{R}|} =P$, as we wanted to see.
\end{example}

\medskip

\begin{example}[Completely depolarising map]
Consider the MPS given by matrices 
\be
&&A^0 = \frac{1}{\sqrt{2}}|0\ra\la 0| , \quad A^1 = \frac{1}{\sqrt{2}}|0\ra\la 1|, \nn\\
&&A^2 = \frac{1}{\sqrt{2}}|1\ra\la 0|, \quad A^3 = \frac{1}{\sqrt{2}}|1\ra\la 1|. 
\ee
The corresponding transfer matrix is the completely depolarising map 
 $\mc{P}(\rho) = \mathrm{tr}(\rho)(I/2)$, which is a projector quantum channel, and thus has a continuum limit. 
Following \cref{thm:thermo}, 
we consider $H=0$ and jump operators 
$R_0=(1/\sqrt{2})|0\ra \la 1| $ and 
$R_1=(1/\sqrt{2})|1\ra \la 0|$, 
which give rise to the following cMPS with open indices on a segment $\mc{R}$ of length $|\mc{R}|=l$, 
\be
\phi_\mc{R} &=& e^{-l/4}[ I 
+\sum_{\alpha=0,1} R_{\alpha} \sum_{N\geq 1, N \textrm{odd}} 
\frac{1}{2^{(N-1)/2}} \times \nonumber\\
&&
\int_{0\leq x_1\leq \cdots x_N\leq l} 
dx_1\cdots dx_N
\hat\psi_\alpha^\dagger(x_1)
\hat\psi_{\alpha+1}^\dagger(x_2) 
\ldots
\hat\psi_\alpha^\dagger(x_N)
 \nonumber\\
&
+&\sum_{\alpha=0,1} R_{\alpha}R_{\alpha+1} \sum_{N\geq 2, N \textrm{even}}
\frac{1}{2^{(N-2)/2}} \times \nonumber\\
&&
\int_{0\leq x_1\leq \ldots x_N\leq l} 
dx_1\ldots dx_N
\hat\psi_\alpha^\dagger(x_1)\hat\psi_{\alpha+1}^\dagger(x_2) \cdots
\hat\psi_{\alpha+1}^\dagger(x_N)
]|\Omega_{\mc{R}}\ra, \nonumber\\
\ee
where the creation operators are of type $\alpha,\alpha+1,\alpha,\alpha+1,\ldots $ ($N$ times), and the sum on $\alpha$ is modulo 2. Note that we have split the sum between $N$ odd and $N$  even, and for this reason the creation operators differ in their final subindex. 
This gives rise to 
\be
E_{|\mc{R}|} = (\phi_\mc{R},\phi_\mc{R}) &=& e^{-l/2}[I\otimes I + 2 \sinh(l/2)\sum_{\alpha=0,1}R_{\alpha} \otimes R_{\alpha} \nonumber \\
&&+ 4 ( \cosh(l/2) -1)\sum_{\alpha=0,1}R_{\alpha}R_{\alpha+1} \otimes R_{\alpha}R_{\alpha+1}]. \nonumber\\
\ee
We again see that $\lim_{|\mc{R}|\to\infty}E_{|\mc{R}|} =P$.
\end{example}

\medskip

\begin{example}[The bracket state] 
\label{ex:bracket}
Consider the transfer matrix $E_a=Pe^{aL}$ where 
\be
P&=& I \otimes I + \sigma_z\otimes \sigma_z, \\ 
\mc{L}(\rho) &=& \frac{\gamma}{a}(\sigma_x \rho \sigma_x -\rho), 
\ee 
(one can verify that $PL=PLP$), and thus 
\be
E_a=Pe^{aL}= I\otimes I + e^{-2\gamma} \sigma_z\otimes \sigma_z . 
\label{eq:Eabracket}
\ee
Note that $\mc{E}_a$ has a non-degenerate eigenvalue 1 (namely $\mc{E}_a(I)=I$), and two 0 eigenvalues   (namely $\mc{E}_a(\sigma_x)=\mc{E}_a(\sigma_y)=0$). 
This channel can be seen as a convex combination of the channel corresponding to the equal superposition of ferromagnetic states, 
$E_a^{(\mathrm{f})} = |00\ra\la 00| +|11\ra\la 11|$, 
 and the one corresponding to the superposition of antiferromagnetic states,
$E_a^{(\mathrm{af})} = |00\ra\la 11| +|11\ra\la 00|$, namely 
\be
E_a = p E_a^{(\mathrm{f})} + q E_a^{(\mathrm{af})},
\ee
where 
\be
p=\frac{1+e^{-2\gamma}}{2}, \quad 
q=\frac{1-e^{-2\gamma}}{2} . 
\label{pqbracket}
\ee

One family of MPS $\mc{V}(A)$ whose transfer matrix is $E_a$ [Eq.\ \eqref{eq:Eabracket}] is given by the matrices
\be
A^0 = \sqrt{p} |0\ra\la 0|, \quad A^1= \sqrt{p}|1\ra\la1|,\nn\\
A^2 = \sqrt{q} |0\ra\la 1|, \quad A^3= \sqrt{q}|1\ra\la 0|, 
\label{bracketAs}
\ee
which result in the state
\be
|V_N(A)\ra =
{\sum_{i_1,\ldots,i_N}}'
p^{(n_0+n_1)/2}
q^{(n_2+n_3)/2}
|i_1,\ldots,i_N\ra ,
\ee
where ${}'$ indicates that the sum is over `allowed' states only, which simply means states with a non-zero coefficient.
The allowed states can be graphically understood as follows: 
we represent the physical index
$0$ by a dot ,$.$, 
$1$ by a dash ,$-$, 
$2$ by an opening bracket ,$($,
and $3$ by a closing bracket ,$)$. 
The allowed states are those such that every bracket which is opened is closed before opening another bracket, 
and which contain zero or more dots outside the brackets, 
and zero or more dashes inside the brackets. 
Examples of allowed states are 
$...(--)....(-)()....$, 
$(-)()()....$,
and $-----$. 
Every bracket contributes with $\sqrt{q}$ (and contains the `antiferromagnetic character' of the state) and every point or dash contributes with $\sqrt{p}$ (and contains the `ferromagnetic character' of the state). 
For this reason we refer to this state as the `bracket state'. 

The continuum limit state of $|V_N(A)\ra$ can be represented by a generalised cMPS $|\Phi_{\mc{R}}[\{v_i\},\{B_i\},Q,\{R_\alpha\}]\ra$ with 
$\mc{R}=[0,Na]$, 
 with $|v_i\ra =|i\ra$, the computational basis, $B_0=\ket{0}\bra{0}$, $B_1=\ket{1}\bra{1}$, $R=\sqrt{\gamma'}\sigma_x$, with $\gamma':=\gamma/a$, namely 
\be
&&|\Phi_{\mc{R}}[\{v_i\},\{B_i\},Q,\{R_\alpha\}]\ra = \nn\\
&& = \ket{0}\otimes\mathrm{tr}_{\mathrm{aux}}\Bigg\{\ket{0}\bra{0}\mathcal{T}\!\exp\Bigg[\int_{\mc{R}} -\frac{\gamma' I}{2}\otimes I\nn\\
&&+\sqrt{\gamma'}\sigma_x\otimes\hat\psi^{\dagger}(x)dx\Bigg]\Bigg\}\ket{\Omega_{\mc{R}}}\nn\\
&&+\ket{1}\otimes\mathrm{tr}_{\mathrm{aux}}\Bigg\{\ket{1}\bra{1}\mathcal{T}\!\exp\Bigg[\int_{\mc{R}} -\frac{\gamma' I}{2}\otimes I\nn\\
&&+\sqrt{\gamma'}\sigma_x\otimes\hat\psi^{\dagger}(x)dx\Bigg]\Bigg\}\ket{\Omega_{\mc{R}}}.
\label{eq:contbracket}
\ee
In addition, according to \cref{thm:thermo}, $\mc{P}$ can be obtained as $\mc{P} = \lim_{t\to \infty} e^{t\tilde{\mc{L}}} $, 
where 
\be
\tilde{\mc{L}}(\rho) = \sigma_z \rho \sigma_z -\rho.
\ee
Note that in this case $\mc{L}$ and $\tilde{\mc{L}}$ commute, $\mc{L}\tilde{\mc{L}}=\tilde{\mc{L}}\mc{L}$. 
\end{example}


The bracket state bears some similarity to the AKLT state \cite{Af88}, 
whose MPS representation is given by
\be
A^0=\frac{1}{\sqrt{3}}\sigma_z,\;\; A^+=\sqrt{\frac{2}{3}}\ket{1}\bra{0},\;\; A^-=-\sqrt{\frac{2}{3}}\ket{0}\bra{1}.
\ee
Identifying the physical indices  
$0$ with a dot ,$.$, 
$+$ with an opening bracket ,$($,
and $-$ by a closing bracket ,$)$, we see that 
the `allowed' states are those such that every bracket which is opened is closed before opening another bracket, 
and which contain zero or more dots inside or outside the brackets. 

However, the AKLT state is obviously not a particular case of the bracket state, since the physical dimension of the former is $3$ whereas that of the latter is 4. 
Their transfer matrices are also very different: 
the transfer matrix of the AKLT state is $E_{\rm AKLT}= \text{diag} (1, -1/3, -1/3, -1/3)$ in the Pauli basis, i.e.\ it has no projector $P$. 
In fact, $E_{\rm AKLT}^2 =e^L$, and thus the AKLT has a \emph{coarse} continuum limit \cite{De17b}.

\section{Conclusions and outlook}
\label{sec:conclusions}

In summary, we have proposed a generalised ansatz of continuous MPS which can express the continuum limit of any MPS (\cref{thm:main}), according to the definition of continuum limit given in \cref{def:cl}. 
The ansatz consists of a sum of cMPS, where each cMPS has a different boundary operator 
and is attached to an ancilla state. 
The boundary operators are given by the Kraus operators of the projector quantum channel $P$. 
We have shown that this ansatz can be interpreted as the concatenation of an element which is at the closure of the set of cMPS (which can be thought of the thermodynamic limit of a cMPS, or a cMPS with matrices of unbounded strength), and a cMPS (\cref{thm:thermo}). 

This work leaves several open questions.
As mentioned in \cref{ssec:discussion}, there is, on the one hand, some freedom in the choice of the generalised cMPS, 
and, on the other hand, our ansatz does not explicitly  impose  the fact that $PL=PLP$. 
That is, in the definition of a generalised cMPS, we need to additionally demand that the matrices $\{B_i\},Q, \{R_\alpha\}$ satisfy $PL=PLP$. 
There might exist another ansatz in which this condition is built-in in the structure of the state; 
this would obviously also reduce some of the freedom in the choice of the generalised cMPS. 
A good place to seek inspiration may be the situation for $G$-injective MPS \cite{Sc10b}. 
These have a parent hamiltonian whose degeneracy is given by the number of conjugacy classes of $G$, and 
the ground state subspace of the parent hamiltonian is obtained precisely by considering superpositions of MPS with different boundary conditions, which is reminiscent of our ansatz. 
Whether the two ideas are in fact connected is a matter of future work. 
Solving this question may also allow us to construct  a physical hamiltonian defined directly at the continuum whose ground state is a generalised cMPS.

Another important question is what is the physical nature of the ancilla Hilbert space $\mathbb{C}^K$ in the ansatz of generalised cMPS, that is, to understand  what observables give access to this ancilla space.

This work could be extended to the non-homogeneous case. 
From the perspective of the transfer matrix, the natural non-homogeneous generalisation of an infinitely divisible channel would be a channel of the form $E = P \,\mc{T} \!\exp (\int  dx L(x))$, where $P^2=P$ and $PL(x) = PL(x) P$ for all $x$. 
(These quantum channels have not been studied, to the best of our knowledge.) 
One could propose a non-homogeneous generalised cMPS ansatz, with the criterion that their transfer matrix precisely corresponds to this channel. 
In this case, however, one would additionally need to characterise the set of (translationally invariant or not) MPS which have a continuum limit, for a new definition of the latter (which may involve a non-homogeneous refining procedure, and some other regularity condition in the limit).

Another open question is whether our framework can be used to describe dynamics in combination with the time-dependent variational principle (TDVP) \cite{Ha14b}. 
This would require, first, the extension of the framework presented in \cite{Ha14b} to cMPS, 
for which one could use the study of tangent vectors of cMPS given in \cite{Ha13}.
Then, this framework could be extended to generalised cMPS. 
For the latter,  the freedom in the choice of parameters (discussed in \cref{ssec:discussion}) will be very important, 
as well as the fact that the condition $PL=PLP$ is not yet built-in in the ansatz.

\emph{Acknowledgements.} We thank 
I.\ Cirac, 
A.\ L\"auchli, 
T.\ Osborne, 
D.\ P\'erez-Garc\'ia, and
N.\ Schuch for discussions. 
We thank the referees for the careful reading of our manuscript and their comments to improve the paper.
MBJ acknowledges F. Mivehvar for discussions  and the support from the ``Doktoratsstipendium aus der Nachwuchsf\"orderung" of the University of Innsbruck.

\appendix

\section{Proof of \cref{thm:thermo}}
\label{app:thermo}

In this Appendix we prove \cref{thm:thermo}. We start by restating it.

\begin{proposition-nn}[\cref{thm:thermo}]
Let $\mc{P}$ be the projector quantum channel given by \eqref{eq:P1}. 
Then the Liouvillian of Lindblad form $\tilde{\mc{L}}$ given by \eqref{eq:Cthermo} is such that 
\be
\mc{P} =\lim_{t\to \infty}e^{t \tilde{\mc{L}} [Q,\{R_{k,1},R_{k,2}\}_{k=1}^n]}. 
\ee 
\end{proposition-nn}

\begin{proof}
Throughout the proof we will write $\mc{L}$ instead of $\tilde{\mc{L}}$ in order to simplify the notation. 
We will prove the statement for the following cases: 
\begin{itemize}
\item[(i)] The completely depolarising map, which has $n = 1$, $D_{1} = 1$ and $m_1= D$. 
\item[(ii)] Case $n=1$ (and $D_{1}$ and $m_1$ unfixed). 
\item[(iii)] Case $D_{k}=1$ for all $k$ (and $n$ and $m_{k}$ unfixed).
\end{itemize}

(i) First consider the case $n=1,D_1=1$, so that $\pp(\rho)=\mathrm{tr}(\rho)\sigma$, with $\sigma=\sum_{i=1}^{D}\theta_i^2\ket{v_i}\bra{v_i} >0$ and $\mathrm{tr}(\sigma)=1$. We claim that this can be written as $\pp=\lim_{t\to\infty}e^{t\L[Q,R_1,R_2]}$ with $H=0$,
\be
R_1=\sum_{i=1}^{D-1}\theta_i\ket{v_i}\bra{v_{i+1}},\quad
R_2=\sum_{i=1}^{D-1}\theta_{i+1}\ket{v_{i+1}}\bra{v_i}. 
\ee
We will show that $\sigma$ is the only fixed point of $e^{t\L}$, since, by \cite[Proposition 7.5]{Wo11}, this implies that the map is primitive (i.e.\ $e^\mc{L}$ does not have any other eigenvalue of modulus 1). 
First note that $\L(\sigma)=0$, so that we only have to see that $\L$ has no other 0 eigenvalue. 
Observe that, for any $\rho$ which is orthogonal to $\sigma$, i.e., $\mathrm{tr}(\sigma^{\dagger}\rho)=0$, we have that $\L(\rho)\neq 0$, as can be easily seen using the form of $\mc{L}$. 
If $\rho$ has only off-diagonal terms, i.e., $\rho=\sum_{k,l=1}^{D}\alpha_{k,l}\ket{v_k}\bra{v_l}$ for $k\neq l$, then it is immediate to see that $\mc{L}(\rho)\neq0$. 
If $\rho$ is diagonal but different from $\sigma$, i.e., $\rho=\sum_{k=1}^{D}\alpha_k\ket{v_k}\bra{v_k}$, then $\L\left(\rho\right)\neq 0$. 
Finally, if $\rho$ is a combination of diagonal and off-diagonal terms, $\rho=\sum_{k,l=1}^{D}\alpha_{k,l}\ket{v_k}\bra{v_l}+\beta_k\ket{v_k}\bra{v_k}$, it can be easily verified that $\L(\rho)\neq0$ as well.

(ii) Now consider the case $n=1$, so that $\M_D = \M_{D_1}\otimes \M_{m_1}$, so that the projector quantum channel is $\pp(\rho^{(1)}\otimes\rho^{(2)})=\text{id}(\rho^{(1)})\otimes \mathrm{tr}(\rho^{(2)})\sigma$ with $\sigma=\sum_{i=1}^{m_1}\theta_i^2\ket{v_i}\bra{v_i}>0$ and $\mathrm{tr}(\sigma)=1$. It is straightforward to see that this can be written as $\pp=\lim_{t\to\infty}e^{t\L[Q,R_1,R_2]}$ with $H=0$, 
\be
R_1=I_{D_1}\otimes\sum_{i=1}^{m_1-1}\theta_i\ket{v_i}\bra{v_{i+1}}, \\
R_2=I _{D_1}\otimes\sum_{i=1}^{m_1-1}\theta_{i+1}\ket{v_{i+1}}\bra{v_i}. 
\ee

(iii) Now consider the $D_{k}=1$ for all $k$, so that $\pp(\rho)=\bigoplus_{k=1}^{n}\mathrm{tr}(\rho_k)\sigma_k$. It can be written as $\pp=\lim_{t\to\infty}e^{t\L [Q,\{R_\alpha\}]} $ with 
\be
\label{c3}
R_{k,1}=\ket{k}\bra{k}\otimes\Big(\sum_{i=1}^{m_{k}-1}\theta_{k,i}\ket{v_{k,i}}\bra{v_{k,i+1}}\Big), \\
R_{k,2}=\ket{k}\bra{k}\otimes\Big(\sum_{i=1}^{m_{k}-1}\theta_{k,i+1}\ket{v_{k,i+1}}\bra{v_{k,i}}\Big).
\ee
where $k=1,\dots,n$.

Finally, putting these three building blocks together, 
it is immediate to see that the statement holds for a generic $\mc{P}$ with unfixed $n,D_k$ and $m_k$. 
\end{proof}

\section{Characterisation of $PL=PLP$ for a special case}
\label{app:PLP}

In this Appendix we characterise the implications of the condition $\mc{P}\mc{L} =\mc{P}\mc{L} \mc{P}$ for a special case of $\mc{P}$ and $\mc{L}$. 
Namely, we consider a special case of the projector quantum channel $\mc{P}:\mc{M}_D \to \mc{M}_D$ given by \eqref{eq:P1} in which $D_0=0$, $D_k=D_1$ and $m_k=m_1$ for all $k>0$, so that $\mc{M}_D = \bigoplus_{k=1}^n (\mc{M}_{D_1}\otimes \mc{M}_{m_1})$, which is equivalent to $I_n \otimes \mc{M}_{D_1}\otimes \mc{M}_{m_1} $.

In addition, we assume that $\mc{L}$ has a single jump operator $R$, 
with the tensor product structure
\be
R=S\otimes T \otimes V, 
\ee 
with $S\in \mc{M}_n$, 
 $T\in \mc{M}_{D_1}$ and 
 $V\in \mc{M}_{m_1}$. 
Similarly, we assume that the Hamiltonian $H$ consists of a single term with the same structure, 
\be
 H=A\otimes B \otimes C, 
\ee
 with 
 $A\in \mc{M}_n$, 
 $B\in \mc{M}_{D_1}$ and 
 $C\in \mc{M}_{m_1}$.

\begin{proposition}\label{thm:PL=PLP}
Consider a projector quantum channel $\mc{P}:\mc{M}_D \to \mc{M}_D$ given by \eqref{eq:P1}, and $\mc{L}[Q,R]$ given by \eqref{eq:L} with 
$H=A\otimes B \otimes C$ and $R=S\otimes T \otimes V$.
If $\mc{P}\mc{L}=\mc{P}\mc{L}\mc{P}$, then
\begin{enumerate}[label=(\roman*),ref=(\roman*)]
\item \label{im:i}
$A$ is diagonal and either $B\propto I$ or $C\propto I$. 
\end{enumerate}
For $R$, either
\begin{enumerate}[label=(\alph*)]
\item \label{im:a}
$S$ satisfies
\begin{subequations}
\be
S_{k,l}\bar S_{k,m} &= &0, \label{eq:pinching1}\\
S_{k,k} \bar S_{k,l} &= & S_{l,k} \bar S_{l,l},\label{eq:pinching2}
\ee
\label{eq:pinchingmapconditions}
\end{subequations}
$T\propto I$ and $V\propto U$, or
\item \label{im:b}
$S$ has one non-zero element per row, $T\not\propto I$ and $V\propto U$, or
\item \label{im:c}
$S$ is diagonal, $T\propto I$ and $V\not\propto U$.
\end{enumerate}
where $U$ is a unitary.
\end{proposition}

Recall that $I$ denotes the identity matrix. 

If $n$, $D_1$ or $m_1$ are 1, then the only cases that hold are the ones in which the corresponding condition is trivial. That is, 
if $n=1$, then all three cases are possible. 
If $D_1=1$, then case \ref{im:a} and \ref{im:c} are possible. 
If $D_1=2$, then case \ref{im:b} and \ref{im:c} are possible. 
If $D_1=m_1=1$ then case \ref{im:a} is possible. 
If $n=D_1=1$ then $V$ must be a unitary. 
If $n=m_1=1$ then $T$ must be the identity. 

Note that Eq.\ \eqref{eq:pinching1} implies that $S$ has at most one non-zero off-diagonal element in every row. 
Examples would be a diagonal $S$, or $S$ being a permutation matrix times a diagonal matrix, or a matrix which is zero everywhere except for a column, e.g.\ $S=\sum_{j=1}^n\ket{j}\bra{1}$. In the latter two cases, $S$ can additionally have non-zero diagonal elements, as long as the symmetry conditions of \eqref{eq:pinching2} are fulfilled.
For example, consider $\rho_1,\rho_2\in \mc{M}_2$, and the projector 
\be
\pp(\rho_1\otimes \rho_2)= 
\Big(|0\ra\la 0|\rho_1 |0\ra \la 0| + 
|1\ra\la 1|\rho_1 |1\ra \la 1| \Big) 
\otimes \mathrm{tr}(\rho_2) \frac{ I }{2}.\nn\\
\ee
Then we could have, for example, $R=\sigma_x\otimes\sigma_z$, corresponding to case \ref{im:a}, or $R=\sigma_z\otimes\ket{0}\bra{1}$ corresponding to case \ref{im:c}.

\begin{proof}
Throughout the proof we will use that $\sigma_k$ is full rank and hence can be inverted, and that $\mathrm{tr}(\sigma_k)=1$. 
We also consider an input state of the form $\rho=\rho_1\otimes \rho_2\otimes \rho_3$. 

For the imaginary part of $PL=PLP$, the component $\la k,k|\dots|l,k\ra$ (with $l\neq k$) of the first tensor product implies that
\be
-i A_{k,l} (\rho_1)_{k,l}B\rho_2 \mathrm{tr}(C\rho_3)=0,
\ee
which implies that $A$ has to be diagonal. 
The rest of the components yield trivial identities, since $B$ and $C$ are Hermitian, thus proving \ref{im:i}.

For the real part of $PL=PLP$, 
the component $\la k,k|\dots|l,m\ra$ of the first tensor product, with $k\neq l$, $k\neq m$ and $m\neq l$ gives
\be
S_{k,l}\bar S_{k,m} (T\rho_2 T^\dagger ) \mathrm{tr}(V\rho_3 V^\dagger)=0, 
\ee
for all $\rho_3$.
This implies \eqref{eq:pinching1}. 
The components 
$\la k,k|\dots|k,k\ra$, 
$\la k,k|\dots|l,k\ra$, and 
$\la k,k|\dots|l,l\ra$ 
of the first tensor product give, respectively, 
\be
&&
\Big\{|S_{k,k}|^2 (T\rho_2 T^\dagger) -
\frac{1}{2}(S^{\dagger}S)_{k,k} [(T^{\dagger}T\rho_2) +(\rho_2 T^{\dagger} T)]\Big\}\times
\nonumber \\
&&\quad\times \Big\{\mathrm{tr}(V\rho_3 V^{\dagger})- \mathrm{tr}(V\sigma_k V^{\dagger}) \mathrm{tr}(\rho_3)\Big\}=0,
\label{eq:condition222} 
\\
&&
\Big\{S_{k,l}\bar S_{k,k} (T\rho_2 T^\dagger ) - \frac{1}{2}(S^{\dagger}S)_{k,l} (T^\dagger T \rho_2)\Big\}
 \mathrm{tr}(V\rho_3 V^\dagger)=0,\qquad
\label{eq:condition22} 
\\
&&
|S_{k,l}|^2 (T\rho_2 T^\dagger) \Big\{\mathrm{tr}(V\rho_3 V^\dagger) -\mathrm{tr}(V\sigma_k V^\dagger)\mathrm{tr}(\rho_3)\Big\} = 0,
\label{eq:condition2}
\ee
for all $\rho_2$ and $\rho_3$.
First note that if $T\not\propto I$ and $V\not\propto U$ then Eqs.\ \eqref{eq:condition222} and Eq.\ \eqref{eq:condition2} imply that $S=0$, which is false by assumption. Hence there are three cases:
\begin{enumerate}[label=(\alph*)]
\item 
If $T\propto I$ and $V\propto U$ then Eq.\ \eqref{eq:condition22} implies that $S_{k,l}\bar S_{k,k}=\bar S_{l,k} S_{l,l}$, which together with \eqref{eq:pinching2} means that $S$ satisfies the conditions of the pinching map, \eqref{eq:pinchingmapconditions}.
\item 
 If $T\not\propto I$ and $V\propto U$ then Eq.\ \eqref{eq:condition22} 
 implies that $S_{k,l}\bar S_{k,k}=\bar S_{l,k} S_{l,l} =0$, i.e.\ $S$ only has one non-zero element per row. 
\item 
If $T\propto I$ and $V\not\propto U$, then Eq.\ \eqref{eq:condition2} implies that $S_{k,l}=0$, i.e.\ $S$ is diagonal.
\end{enumerate}
\end{proof}

\end{document}